\documentclass[a4paper,11pt]{article}
\pdfoutput=1 

\usepackage{jinstpub} 

\usepackage{lineno}
\usepackage[caption=false]{subfig} 
\usepackage{caption}
\usepackage{threeparttable}

\title{\boldmath Allpix Squared Simulations of Multi-element Germanium Detectors for Synchrotron Applications}

\author[a,1]{T. Saleem\note{Corresponding author.}}
\author[a]{F.J. Iguaz}
\author[a]{F. Orsini}

\affiliation[a]{SOLEIL Synchrotron, \\L’Orme des Merisiers, Saint-Aubin BP 48, Gif-sur-Yvette, 91190 France}

\emailAdd{tasneem.saleem@synchrotron-soleil.fr}

\abstract{X-rays spectroscopy experiments at synchrotron facilities were limited for many years by the maximum input-count rate and the signal-to-background ratio of germanium fluorescence detectors. These limitations are related to the germanium semiconductor device, the sensor configuration and its response to the incident X-ray flux at different energies. In order to understand and quantify such limitations, physics simulation of the detector response is a powerful tool to provide guidelines for designing, prototyping and improving detectors, as well as modelling experimental environments, which reduces time and cost of development. For this purpose, a first complete and operational simulation chain based on Allpix Squared framework is presented, customized to multi-element germanium detectors and combined with three-dimensional simulations of the electric field and the weighting potential, based on COMSOL Multiphysics®. Based on this simulation chain, a quantification of charge sharing effect as well as signal-to-background ratio at different beam energies has been made for a germanium detector equipped with and without collimator. In addition, two experimental measurements have been performed on the SAMBA beamline at SOLEIL synchrotron. The experimental data were used to set up the full simulation chain and good agreements have been observed between data and simulation.
}


\begin{document}
\maketitle
\flushbottom

\section{Introduction}
X-rays spectroscopy techniques are commonly used on several beamlines at synchrotron facilities to perform experiments in a large field of sciences such as Physics, Chemistry, Environmental Sciences, Biology, and Surface Material Science. During many years, these experiments have been limited by the sensitivity of fluorescence detectors, where the maximum input count rate, the signal-to-background ratio and the energy resolution are key performance features. With the current and coming upgrade of synchrotron light sources with higher brilliance, it becomes a major challenge to overcome the limitations of current commercial detectors available on the market. For that purpose, the understanding of detector performances and limitations including its associated front-end electronics remains essential.

The objective of this work is to build a full simulation chain adapted to multi-element germanium detectors~\cite{HANSEN1971377,SANGSINGKEOW2003183,Amman:2018oci} to study their response to incident X-rays, in the energy range from 5 to 80~keV, including on-beamline experiments at synchrotron facilities. For realistic modeling of the multi-element germanium detector response, a simulation chain combining three-dimensional electrostatic field simulation using COMSOL Multiphysics®~\cite{comsol} and Allpix Squared framework~\cite{Allpix} is used. It's worth mentioning that Allpix Squared is a generic simulation framework originally developed to study the performance of silicon detectors. This work is the first application of this framework for the simulation of germanium detectors.

After the presentation of a model of multi-element germanium detector in Section~\ref{sec:detmodel} and the simulation flow in Section~\ref{sec:simflow}, including the custom modifications for germanium detectors in X-ray spectroscopy; the simulation chain is applied to a current germanium detector in use at the SAMBA beamline~\cite{SAMBA} of the SOLEIL synchrotron in Section~\ref{sec:chargesharing}. The detector performance at different beam energies is simulated. Then, the validation of the simulation chain is discussed in Section~\ref{sec:validation}, by comparing the predicted performance by simulation with that obtained in two experimental measurements at the beamline.
The conclusions and an outlook of future simulation studies in Section~\ref{sec:con} complete this paper.

\section{Multi-element germanium detector specifications}
\label{sec:detmodel}
\begin{figure}[htb!]
\centering
\subfloat[]{\includegraphics[width=0.57\textwidth]{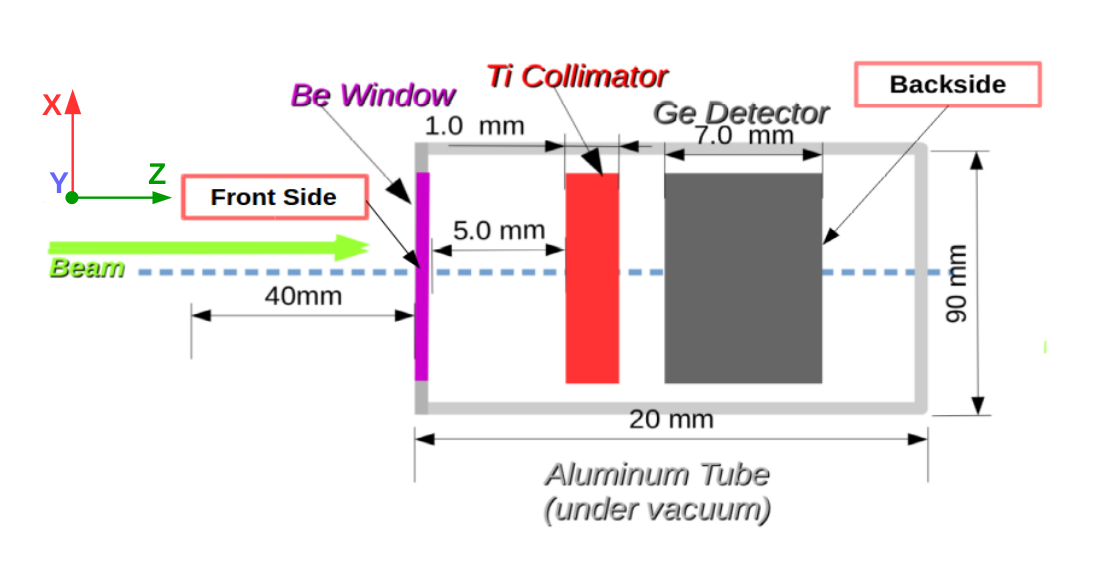}}
\qquad
\subfloat[]{\includegraphics[width=0.37\textwidth]{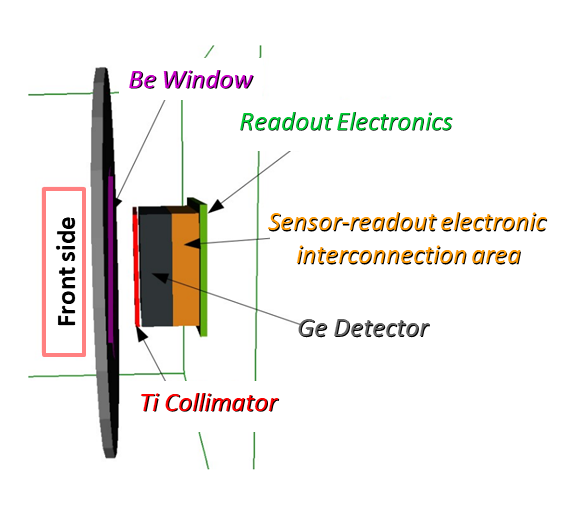}}
\caption{(a) Schematic representation of the multi-element germanium detector under study showing the different simulated components, where the frontside is the one facing the incident X-rays (b) Visualization of the multi-element germanium detector geometry implemented in Allpix Squared framework.}
\label{fig:SAMBAGeometry}
\end{figure}

The multi-element germanium detector used as model for this simulation is a commercial germanium detector (Model: Canberra EGPS 30$\times$30$\times$7-36 PIX). Figure~\ref{fig:SAMBAGeometry} (a) presents a general schematic of the multi-element germanium detector under study and shows the different geometry components. The detector consists of a high-purity germanium (Ge) crystal with an active area of 30~mm~$\times$~30~mm and a thickness of 7~mm. The germanium crystal is located at 5~mm behind a 125~\textmu m thick beryllium (Be) window with a diameter of 46~mm. On the backside, the crystal is segmented into 6 $\times$ 6 pixel contacts, with an implant size of 4.2~mm $\times$ 4.2~mm. The detector under study has a titanium (Ti) collimator with a thickness of 1~mm at a distance of 1~mm in front of the germanium crystal. The collimator has 36 square holes, distributed in a squared pattern like the crystal implants. Each hole has a width of 4.2~mm and a pitch of 5~mm. Collimator holes are aligned with the pixel grid such that titanium covers the area between implants. The detector is surrounded by an aluminium tube of 90~mm inner diameter and 2~mm thick. The aluminium tube keeps the detector under vacuum. The detector operates in electron collection mode at a nominal bias voltage of -800~V.

This model of a multi-element germanium detector has been implemented as detector geometry in Allpix Squared framework, as it is shown in Figure~\ref{fig:SAMBAGeometry} (b). In this figure, the beryllium window is in purple and the frontside of the aluminium tube is in grey. The surrounding tube is invisible for a better visualization of the inner parts of the detector. Inside the aluminium tube, the collimator is in red, the germanium sensor in black, the sensor-readout electronic interconnection area in orange and the readout electronics in green.

\section{Physics processes and simulation flow}
\label{sec:simflow}
In this section, a brief explanation of the physics processes resulting from X-ray interaction inside a multi-element germanium detector, followed by a full description of the simulation flow used in this work are presented.

\begin{figure}[htb!]
\centering
\includegraphics[height = 0.5\textwidth]{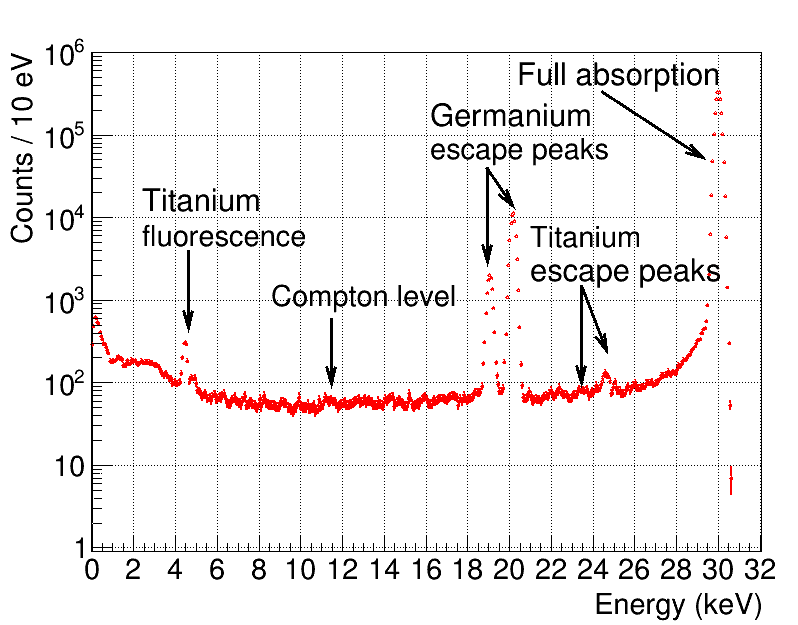} 
\caption{Example of a simulated energy distribution deposited by 30~keV photons and measured by a germanium detector equipped with a titanium collimator. All peaks and regions of interest that correspond to different physics processes are identified. }     
\label{fig:GermaniumPhysics}
\end{figure}

In the energy range from 5 and 100~keV, the photoelectric interaction is the dominant process. In this case, the photo-electron loses most of its energy by ionization but the remission of germanium fluorescence X-rays at energies of 9.88 and 10.98~keV, corresponding to K$_\alpha$ and K$_\beta$ emission lines of germanium, is also possible for energies above 11~keV,
as shown with the example of Figure~\ref{fig:GermaniumPhysics}.
These germanium fluorescence X-rays may be reabsorbed in the germanium crystal, creating another energy deposit, or escape. Events with two well-separated energy deposits will spread over multiple pixels, resulting in charge sharing events called \textit{fluorescence events} (following the name convention of~\cite{BARYLAK2018234}). All energy deposits promote electrons to the conduction band, creating clouds of electrons and holes. By applying a bias voltage across the crystal, charge carriers drift from the frontside to the segmented backside, where they induce signals in the readout electronics. In our case, holes drift to the frontside and electrons to the backside. While drifting, charge clouds widen due to diffusion effects and carriers drifting to the backside may spread and induce signals over multiple pixels, resulting in charge sharing events called \textit{split events}. Charge sharing events (either \textit{fluorescence} or \textit{split events}) reduce the intensity of the full absorption peak and increase the Compton level (referred in Figure~\ref{fig:GermaniumPhysics}), because the initial X-ray energy cannot be fully reconstructed by actual electronics. These two effects respectively translate into a loss of signal efficiency and an increase of background level, i.e., an undesirable reduction of signal-to-background ratio in synchrotron experiments~\cite{Heald:rv5031}. Induced signals are then digitized and finally collected into a full energy spectrum for each pixel by a Digital Pulse Processor (DPP). Coming back to X-ray interactions, the titanium collimator situated at the frontside of the germanium crystal makes also possible the absorption of titanium fluorescence (at energies of 4.51 and 4.93~keV, K$_\alpha$ and K$_\beta$ lines) in the germanium crystal and, with less probability, of Compton scattered photons from titanium collimator. The energy spectrum shown in Figure~\ref{fig:GermaniumPhysics} has been obtained by simulating $2 \times 10^6$~X-rays. In 94.5\% of the cases, the initial energy of 30~keV has been deposited in one pixel (full absorption). While in the remaining cases, energy is either shared with other pixels or lost either by fluorescence or by Compton scattering.

To model previously mentioned physics processes, the simulation flow is divided into five parts:
the X-ray interaction with matter,
the charge carrier drift using a three-dimensional field map,
the signal induction in the segmented backside using a three-dimension weighting potential map,
the signal digitization and the creation of the energy spectrum in the DPP.
All these steps are done through different modules of Allpix Squared framework~\cite{Allpix}, but using a custom-modified version of the release 1.6.0 (Oct. 2020)~\cite{spannagel_simon_2020_4494619}. The list of custom changes is the following:
\begin{enumerate}
 \item In the module \textit{GeometryBuilder}, germanium material has been defined and used to build the crystal. The other geometry elements, described in Section~\ref{sec:detmodel} (beryllium window, titanium collimator and aluminium tube) and the reference sample, described in Section~\ref{sec:expsetup}, have been also implemented in this module.
 \item In the module \textit{DepositionGeant4}, values of electron-hole pair energy and Fano factor for germanium have been set in the configuration file. The physics list for X-ray polarization (\textit{G4EmLivermorePolarizedPhysics}) has been also defined in this module.
 \item In the module \textit{GenericPropagation}, the Jacoboni-Canali mobility model of carriers for silicon has been replaced by that for germanium at a fixed temperature of 77~K~\cite{BRUYNEEL2006764}. We must note that since the release 2.0 of Allpix Squared framework, a custom mobility model for germanium in \textit{Physics Models} can be defined.
\end{enumerate}

The two three-dimensional maps have been externally simulated by COMSOL Multiphysics®, and the specific features of DPP have been implemented in the post-simulation analysis. A list of simulation steps, Allpix Squared modules and specific simulation parameters is provided in Table~\ref{tab:sim}.

\begin{table}
\caption{List of simulation steps, corresponding module(s) in Allpix Squared framework
and specific parameter values. Electron and hole mobility parameter values, corresponding to a Jacoboni-Canali model, are not set by configuration file but implemented in the mobility model. The spectrum formation by DPP is not made in Allpix Squared framework but in post-simulation analysis.}
\label{tab:sim}
 \resizebox{\textwidth}{!}{
\begin{tabular}{l|l|lll}
\hline
 Step & Allpix Squared module & Parameter & Symbol & Value\\
\hline
 Energy deposition & \textit{DepositionGeant4} & Electron-hole pair energy & $W$ & 2.9~eV\\
  & & Fano factor & $F$ & 0.112\\
\hline
 Drift & \textit{ElectricFieldReader} & Model & & mesh\\
       & & File name & & .init file\\
       & \textit{GenericPropagation} & Temperature & T & 77~K\\
       & & Electron mobility & $\upmu^0_e$ & 38536~cm$^2$/V/s\\
       & & - field correction & $E^0_e$ & 53.8~V/mm\\
       & & - temperature correction  & $\beta_e$ & 0.641\\
       & & Hole mobility & $\upmu^0_h$ & 61215~cm$^2$/V/s\\
       & & - field correction   & $E^0_h$ & 18.2~V/mm\\
       & & - temperature correction & $\beta_h$ & 0.662\\
       & & Charge per step & & 100~e/h\\
       & & Integration time & & 200~ns\\
\hline
Transfer & \textit{WeightingPotentialReader} & Model & & mesh\\
         & & File name & & .init file\\
         & \textit{InducedTransfer} & Induction matrix & & 3$\times$3 \\
\hline
Digitization & \textit{DefaultDigitizer} & Electronics noise & \it{ENC} & 31~e\\
             & & Threshold & & 100~e\\
             & & Threshold smearing & & 0~e\\
\hline
Spectrum & - & Dead time & & 1430~ns\\
         & (offline analysis) & Time resolution & & 300~ns\\
\hline
\end{tabular}}
\end{table}

\subsection{Energy deposition: Geant4 simulation}
\label{sec:Geant4}
The interaction of X-rays with the germanium crystal is simulated by the \textit{Geant4}
library~\cite{Agostinelli:2002hh} using the \textit{DepositionGeant4} module. This Allpix Squared module has been adapted to germanium detectors by modifying the base material from silicon to germanium. Polarization effects have been included by explicitly using \textit{G4EmLivermorePolarizedPhysics} class of Geant4 (version 4.10.6), which also uses low energy models based on Livermore data libraries for interactions of photons with matter. These interaction models are accurate for photon energies between 250~eV up to 100~GeV and can be applied down to 100~eV with a reduced accuracy~\cite{Cirrone2010315}. The energy deposits in the germanium sensors are translated into charge carriers with a conversion factor of 2.9~eV per electron-hole pair and a primary charge fluctuation modeled by a Fano factor of 0.112~\cite{ANTMAN1966272}.

In this work, a photon beam has been simulated for two different cases:
\begin{itemize}
\vspace{-0.2cm}
 \item A uniform square shape beam, directed along the positive z-axis, perpendicular to the detector and X-ray energies between 5 and 80~keV. This case is used to directly illuminate the multi-element germanium detector for charge sharing studies, presented in Section~\ref{sec:chargesharing}.
\vspace{-0.2cm}
\item A 2D Gaussian beam, directed along the x-axis, linearly polarized along z-axis, to irradiate a reference sample to study the resulting fluorescence detected by the detector, located perpendicular to the beam. This case, presented in Section~\ref{sec:validation}, is used to replicate the experimental beam conditions and to validate the simulation results.
\end{itemize}

\subsection{Electrostatic field modeling: COMSOL Multiphysics® simulation}
The non-linear variation of the electric field in three dimensions, within the device structure, can affect the behavior and the predicted performance of the detector. Due to that, a precise three-dimensional electric field description is a crucial ingredient for a proper simulation and understanding of detector characteristics. Moreover, combining electrostatic field simulations with Monte Carlo methods enables realistic modeling of the detector response as shown in Section~\ref{sec:chargesharing}.

The COMSOL Multiphysics® software uses a Finite Element Method (FEM) to simulate the electric field map if the practical dimensions and material properties of the device are given. COMSOL Multiphysics® is used in this work to formulate and solve the differential form of Maxwell’s equations together with a set of initial and boundary conditions, and post-process the results to evaluate non-linear electric field components within the simulated device.

\begin{figure}[htb!]
\centering
\includegraphics[width=0.5\textwidth]{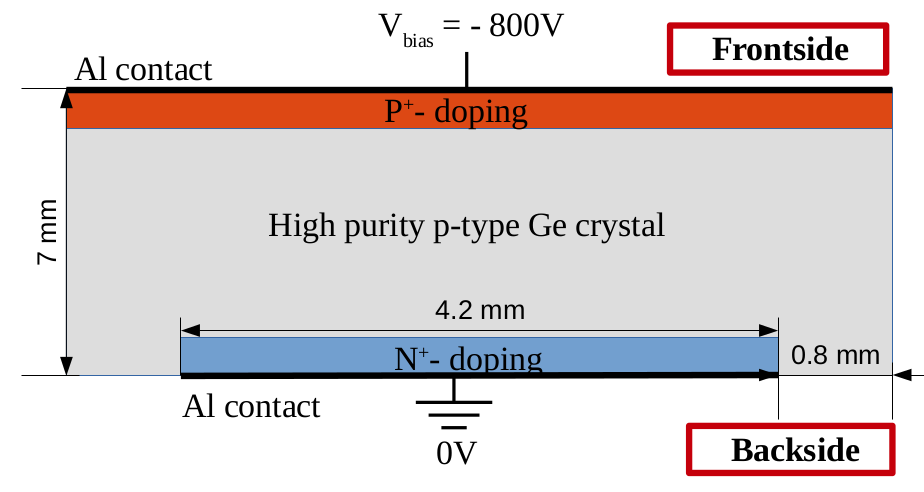}
\caption{Schematic representation of the germanium sensor under study. The high purity germanium crystal is in gray. The N$^{+}$- doped collection electrode (pixel region) of thickness 5~\textmu m  is in blue. The P$^{+}$- doped region with a higher doping concentration of thickness of 10~\textmu m is in red. A bias voltage of -800~V  is applied at the frontside, and the pixel electrode is grounded.}
\label{fig:SAMBA_Schematic_COMSOL}
\end{figure} 

The workflow is as follows: first, the geometry is defined and parametrized; materials are assigned to the different volumes of the model; a semiconductor interface module~\cite{comsol_Semiconductor-module} is chosen with the module’s respective boundary and initial conditions; the element mesh is created; the non-linear solver is selected and subsequently, the results are computed. The phenomenon of avalanche breakdown in germanium sensor has been simulated under the basis of the Okuto-Crowell Impact Ionization model~\cite{OkutoCrowell}. This model requires eight parameters that are specific for each semiconductor material. While for silicon detectors these values are well known, for germanium detectors they are rarely cited in the literature; the ones used for this work can be found here~\cite{OkutoCrowell}.

For these simulations, the detector geometry previously introduced in Section~\ref{sec:detmodel} has been used. A more detailed schematic representation of the germanium sensor only is shown in Figure~\ref{fig:SAMBA_Schematic_COMSOL}. The frontside (facing the incident X-rays) is boron implanted (p$^{+}$- doped electrode) while the backside (segmented side) is phosphorus implanted (n$^{+}$- doped). The inter-pixel spacing is equal to 800~\textmu m. An aluminium contact of 300~nm thickness has been evaporated on the implants (at pixel surface) and on the frontside of the device. The corresponding voltages applied to the different collection electrodes are also shown in Figure~\ref{fig:SAMBA_Schematic_COMSOL}. The detector works in electron collection mode.

The information about the different implanted regions is not available by the manufacturer since this type of information is usually confidential. Therefore, typical values for p-type high purity germanium crystal ($2\times10^{10}$ atom/cm$^3$) and standard dopant concentration for implanted regions have been used: $9\times10^{17}$ atom/cm$^3$ for boron and $2\times10^{20}$ atom/cm$^3$ for phosphorus. Implanted region doping follows a Gaussian profile, with the maximum concentration at the edge of the region and a gradual doping decrease with depth. The simulation includes a single pixel cell with periodic boundary conditions, which allow the field map to be replicated over the entire sensor. For computing time and memory capability reasons, only 1~mm thick volume near pixel implants has been simulated. It has been verified that this volume is sufficiently representative of the non-linear part of the electric field close to the pixel implants, and the electric field has been linearly extrapolated for the remaining depth.

\begin{figure}[htb!]
\centering
\subfloat[]{\includegraphics[width=0.4\textwidth]{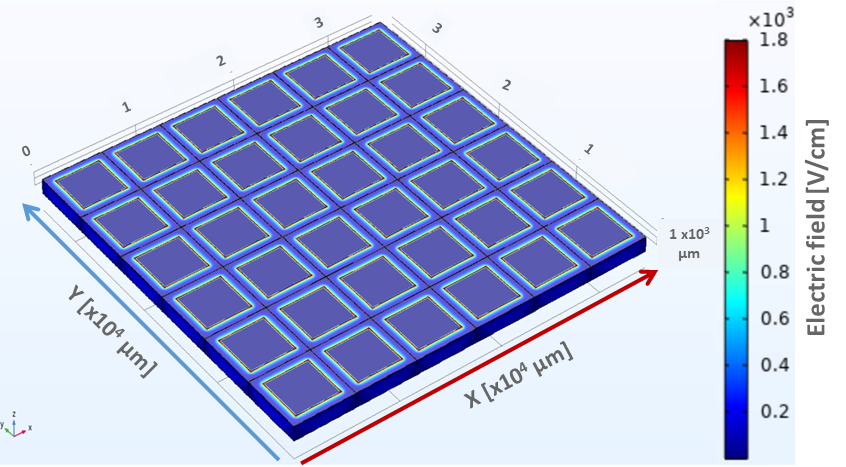}}
\qquad
\subfloat[]{\includegraphics[width=0.5\textwidth]{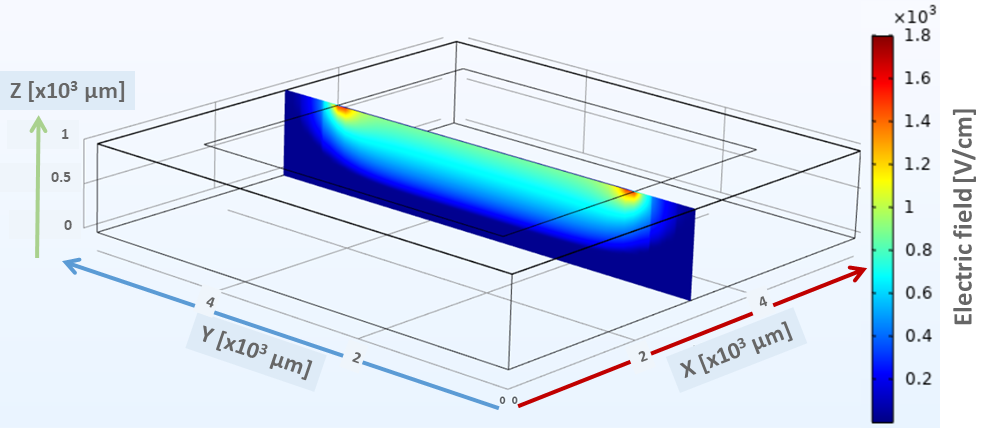}}
\caption{(a) Visualization in the three-dimensions of the magnitude of the electric field [V/cm], throughout the volume of the germanium sensor under study, simulated using COMSOL Multiphysics®. (b) Cross-section through the sensor volume of the same magnitude for one pixel cell. Vertical axis in both figures are inverted with respect to Figure~\ref{fig:SAMBA_Schematic_COMSOL}}
\label{fig:ElectricField}
\end{figure}

A visualization of the magnitude of the electric field in the three-dimensional model is shown in Figure~\ref{fig:ElectricField} (a). A low electric field is present in the inter-pixel region as well as the central pixel region, as indicated by the blue region on the surface of the simulated pixel cell. In contrast, a high-field region is evolving around the pixel edge region, which is the boundary between N$^+$-implant region (blue element in Figure~\ref{fig:SAMBA_Schematic_COMSOL}) and the bulk region (grey element in the same Figure). A cut through the collection electrode, perpendicular to the sensor surface is shown in Figure~\ref{fig:ElectricField} (b). The electric field strength is higher close to the pn-junction around the collection electrode and decreases rapidly towards the sensor frontside.

\begin{figure}[htb!]
\centering
\subfloat[]{\includegraphics[width=0.45\textwidth]{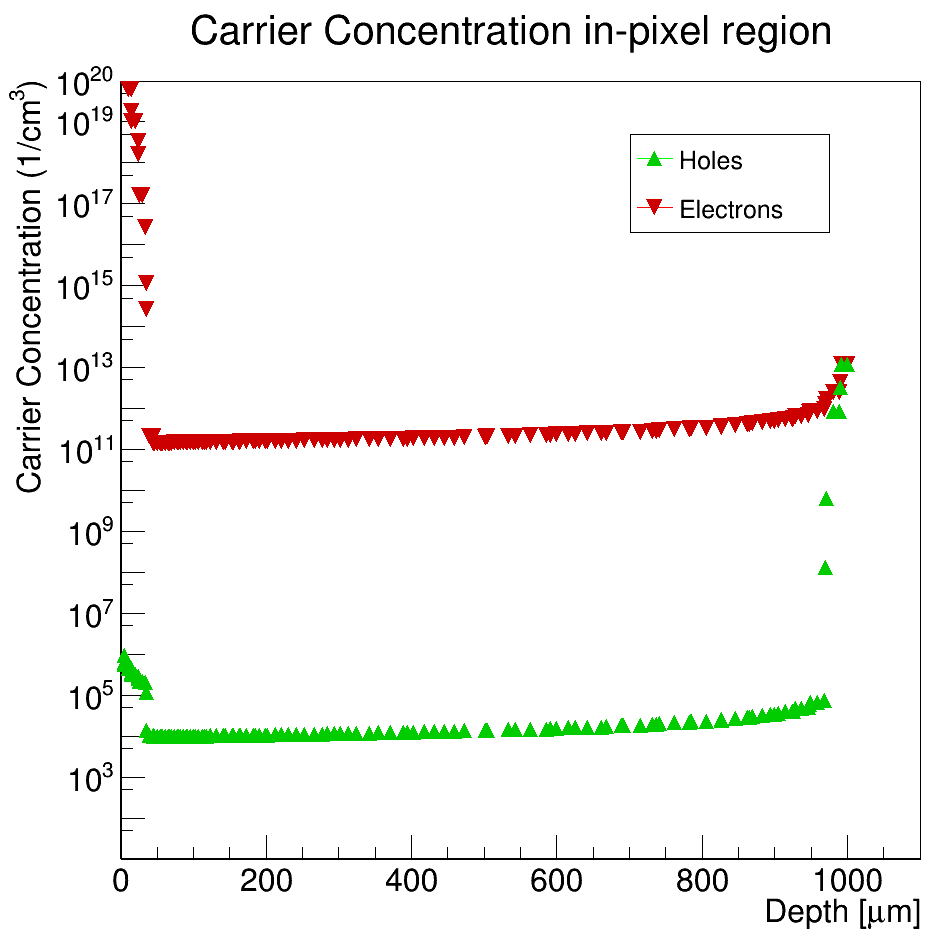}}
\qquad 
\subfloat[]{\includegraphics[width=0.45\textwidth]{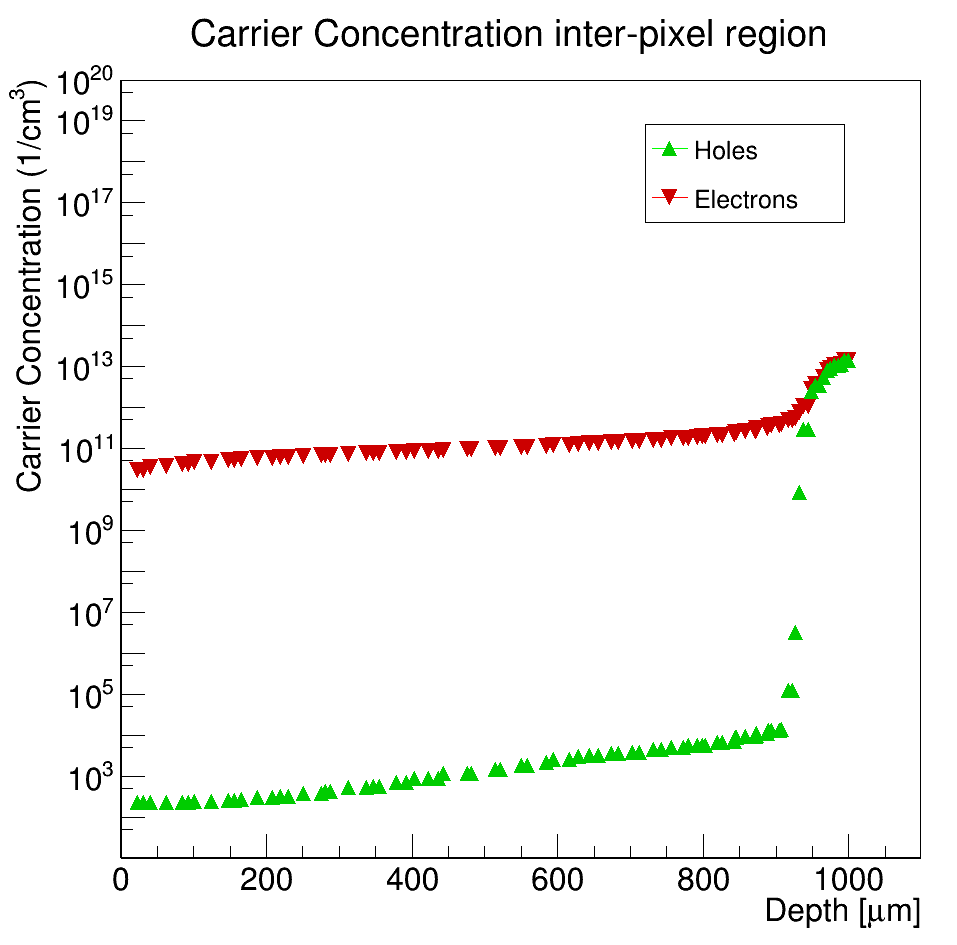}}
\caption{Charge carrier concentration as a function of depth for the in-pixel region (a) and inter-pixel region (b). In both plots, the axis reference is situated at the segmented side (pixels), while the maximum depth is situated at the front implanted side.}
\label{fig:ChargeCarrier}
\end{figure}

Charge carrier concentration, as a function of the sensor depth, is shown in Figure~\ref{fig:ChargeCarrier}, for in-pixel (inside the pixel) region (a) and inter-pixel (between pixels) region (b). Inside the pixel, the charge carrier distribution shows that electron concentration is higher at the backside implant (i.e., electrons are collected by the backside electrode). On the other hand, the hole concentration is lower at the side of the pixel, which is reflecting the fact that the sensor is reversed bias and working in an electron-collecting mode as foreseen by design.

\begin{figure}[htb!]
\centering
\subfloat[]{\includegraphics[width=0.5\textwidth]{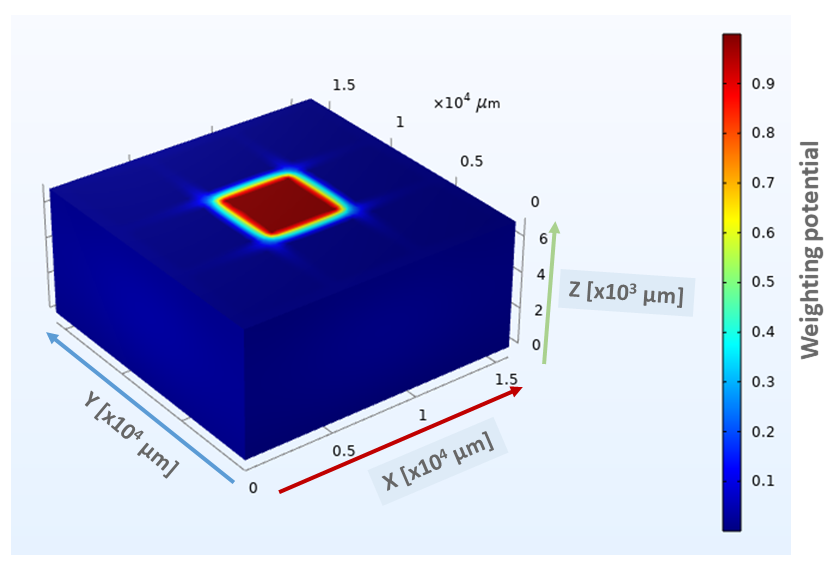}}
\qquad
\subfloat[]{\includegraphics[width=0.43\textwidth]{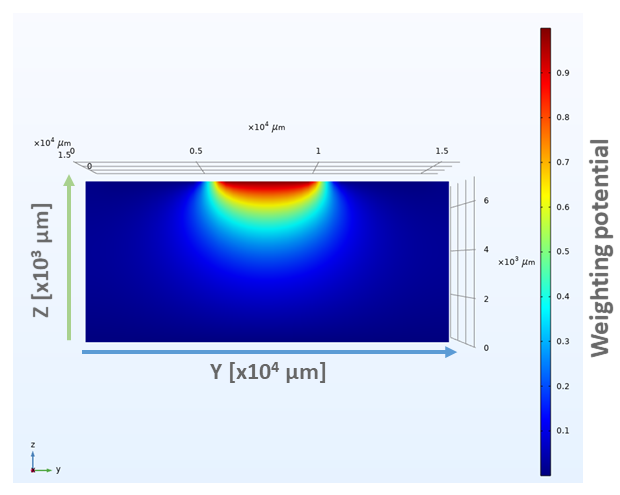}}
\caption{(a) Three-dimensional map of the simulated weighting potentials for 3$\times$3 units cells of the germanium sensor under study and (b) two-dimensional weighting potential slice in a central plane perpendicular to the detector surface. }
\label{fig:WP}
\end{figure}

A second simulation in COMSOL Multiphysics® has been made to create a three-dimensional map of weighting potential in a region of 3$\times$3 units cells. This second map is later imported by Allpix Squared framework, to simulate the induced charge on the electrode resulting from charge carriers moving near electrodes. The weighting potential is a unit-less potential, calculated by setting the central electrode to 1~V and all others to 0~V in a geometry with no dopant concentration (i.e., it is a solution of Laplace's equation). In this case, the whole sensor thickness (7~mm) has been simulated. The resulting weighting potential map is shown in Figure~\ref{fig:WP}.

Before importing the electric field and weighting potential maps into Allpix Squared framework, two steps have been done. In the first step, an output grid (.grd) file containing the adaptive mesh and the evaluated electric field vector or weighting potential has been generated directly within the COMSOL interface. In the second step, the adaptive COMSOL mesh has been interpolated and transformed into a regular grid with a configurable size by a custom script. In the case of the electric field map, an extrapolation by a linear electric field map over the remaining depth up to complete the 7~mm thickness has been made, too. The resulting .init files are imported by the corresponding modules of Allpix Squared framework.

\subsection{Charge carrier drift and signal induction}
\label{sec:Drift}
Charge carriers are propagated along the electric field lines through the germanium crystal until they reach the front and backsides by the \textit{GenericPropagation} module. This module uses a Runge-Kutta integration method~\cite{Runge1895,Kutta} and the mobility is calculated at each step from the interpolated vector obtained from the three dimensional electric field map, imported by the \textit{ElectricFieldReader} module.
Random diffusion is added to every integration step of the charge carrier motion
using Fick's second law and Einstein relation.
The drift mobility parametrization in germanium at 77~K is formally the same as the Jacoboni-Canali model in silicon, but the parameter values have been modified to those measured
along the principal crystallographic axis $\langle 1 0 0\rangle$ for electric fields lower than 3000~V/cm~\cite{BRUYNEEL2006764}. The drift velocity anisotropy, observed in relatively thick (i.e. 7 cm) germanium detectors~\cite{Abt:2010ax}, is small for thin sensors like in our case and has not been included in the simulation. Charge carriers are propagated during 200~ns, the so-called \textit{integration time}. The integration time has been chosen twice the drift time (less than 100~ns), long enough so that charge carriers are collected either to the electrodes or any other germanium surface. Charge carriers are propagated in groups of 100 per step and the integration time per step has been to 1~ns. These two changes, made to speed up the simulation process, do not affect the precision of simulation results in terms of energy spectrum shape and charge carrier arrival time distribution, as it has been verified. Charge carrier recombination has not been included in this simulation step because recombination lifetime values (greater than 10~\textmu s~\cite{PhysRev.87.387}) are longer than the simulated drift time.

The \textit{InducedTransfer} module is used in this work to simulate the induced signal on pixels resulting from charge carriers movement. This module uses the Shockley-Ramo's Theorem~\cite{doi:10.1063/1.1710367,1686997} and a three-dimensional map of weighting potential in multiples of a pixel cell (in our case, 3$\times$3 and generated by COMSOL), imported by \textit{WeightingPotentialReader} module.


\subsection{Signal digitization and energy spectrum}
\label{sec:Digital}
Induced signals on pixels are transformed into a digital signal using the \textit{DefaultDigitizer} module. The applied threshold to each pixel has been fixed to 100~$\text{e}^{-}$. The electronics noise is modeled by a Gaussian distribution centered at zero with a standard deviation of 31~$\text{e}^{-}$. This value for electronics noise is equivalent to an energy resolution (FWHM) of 235~eV at an X-ray energy of 5.9~keV for a DPP using a trapezoidal filtering and an integration time (or peaking time) of 1~\textmu s. This value has been experimentally measured and has been cross-checked by a detector noise model~\cite{Bordessoule2020}.

While the main DPP features, i.e. dead time and time resolution, are not yet modeled in Allpix Squared framework, these features have been modeled in an offline analysis.
The dead time is the mean time for a DPP to digitize the signal amplitude. The time resolution models pile-up effect, i.e., the efficiency to resolve two induced signals close in time. These two parameters are basically defined by the type of digital filter~\cite{Bordessoule2019}.
Values of 1430~ns for measurement time and of 300~ns for time resolution have been set based on experimental data of a germanium detector equipped with XIA-DXP-xMAP DPP~\cite{XIAXMAP,Hubbard1996}.

Simulation results are stored in ROOT~\cite{Brun:1997pa} trees using the \textit{ROOTObjectWriter} module. For each simulated event, the information of the charge carriers (the initial creation in \textit{DepositedCharge} and after diffusion in \textit{PropagatedCharge}) and the fired pixels (\textit{PixelHit}) is stored. The number of absolute simulated X-rays varies between $5 \times 10^7$~photons per value of energy in the case of the direct beam, equivalent to an exposure time of around 140~s at a photon flux of $3.6 \times 10^5$~ph/sec; to $2 \times 10^{11}$~photons in the case of sample irradiation, equivalent to an exposure time of around 6~sec for a photon flux of $3.47 \times 10^{10}$~ph/sec.

\section{Simulation of multi-element germanium detector performance}
\label{sec:chargesharing}
The performance of a semiconductor particle detector is strongly affected by the charge transport properties of the sensor material. Diffusion, in particular, and drift of electrical charges is a main concern in the case of pixelated detectors. A part of the charge created in the sensor volume may be shared between neighboring pixels as explained in Section~\ref{sec:simflow}. This effect is undesirable in radiation detectors since it leads to the underestimation of the energy of incident photons, and as a result, it reduces the signal efficiency. Moreover, the presence of charge sharing effect can seriously degrade the spectral resolution.

In this section, the charge sharing effect in the germanium detector under study is presented, where all results shown have been obtained using the simulation chain presented earlier in this paper. 


Moreover, current detectors in synchrotron experiments, e.g. Silicon Drift Detectors (SDD) and germanium detectors, use an internal collimator to avoid inter-pixel region events and to remove any charge sharing effect. On one hand, using the collimator can be seen as a simple and efficient solution to improve the signal-to-background ratio. On the other hand, using it implies a significant reduction of the detector active area, especially for photons with small incident angles. In the future, synchrotron light sources will produce more intense photon beams and detectors with a smaller pixel size will be required. However, from a technical point of view, it could be difficult to have collimators that fit smaller pixel sizes and the resulting dead area could become a serious limitation of granular detectors. For these reasons, it is interesting to compare performance of germanium detectors with or without using a collimator.

\subsection{Quantification of the charge sharing effect with and without collimator}
In this section, the charge sharing effect is evaluated using two observables: the cluster size and the seed signal. A cluster is defined as a set of pixels that collect charges when one photon is interacting with the detector. Hence, the \textit{cluster size} is equal to the number of pixels belonging to a cluster. The \textit{seed signal} is defined as the ratio between the charge of the pixel in which the incident particle deposited its energy and the total collected charge. In this study, a squared-shaped beam (30~mm $\times$ 30~mm size) of X-rays (30~keV energy), emitted at 5~cm from the detector window and perpendicular to it (along z-axis in Allpix Squared framework reference) has been simulated. The total number of simulated X-rays is 5 $\times$ 10$^7$.

\begin{figure}[htb!]
\centering
\subfloat[]{\includegraphics[width=0.40\textwidth]{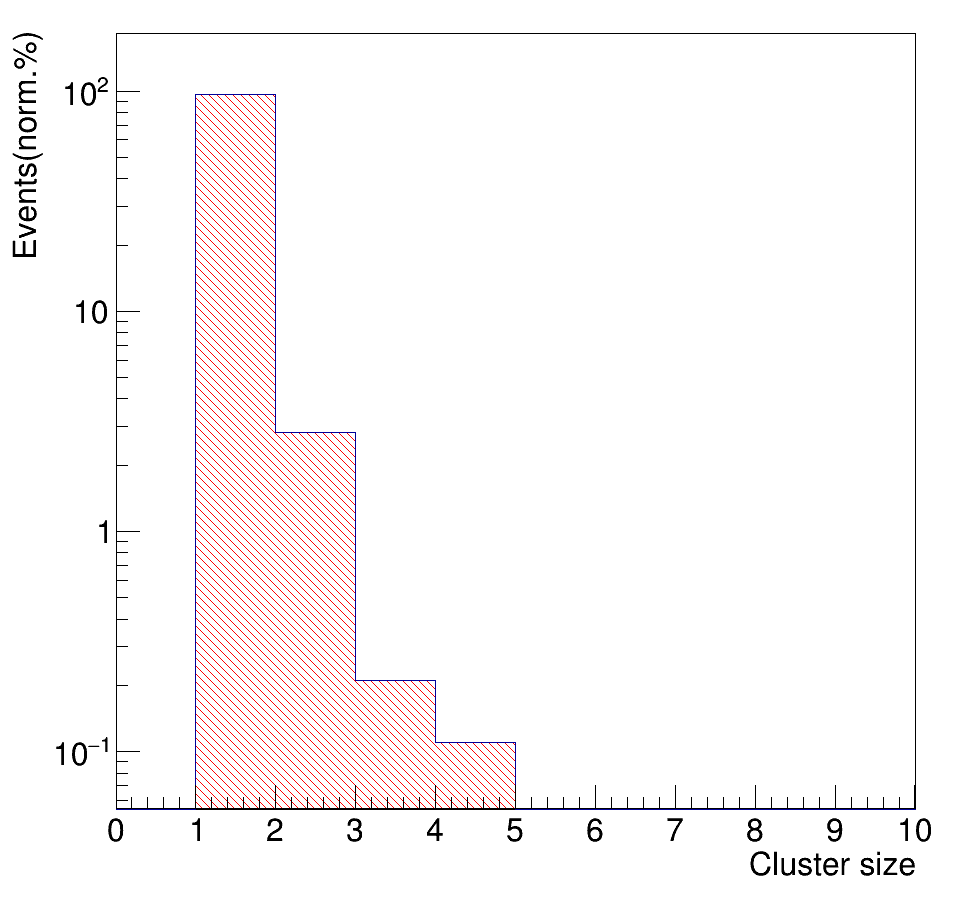}}
\qquad
\subfloat[]{\includegraphics[width=0.44\textwidth]{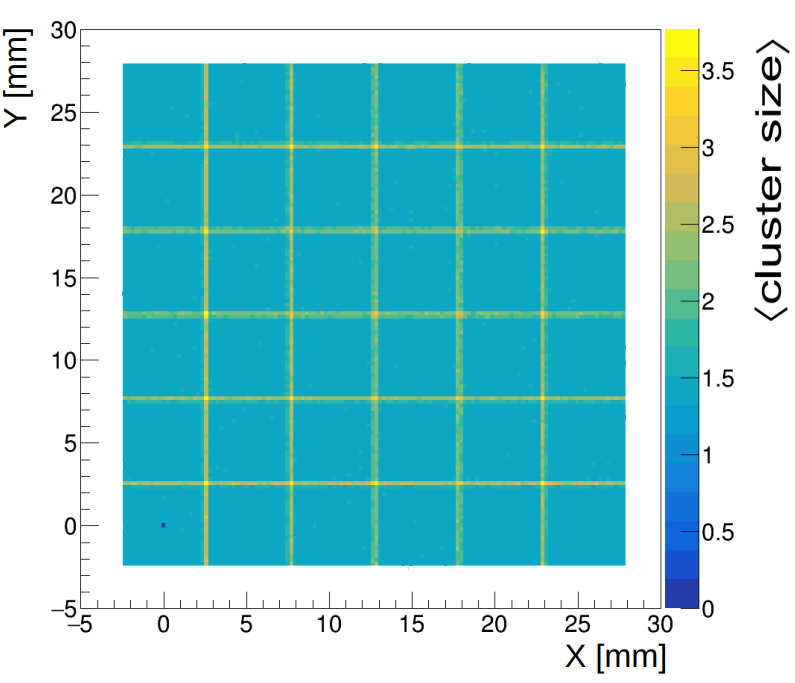}}
\caption{Cluster size distribution (a) and cluster size map (b) for the multi-element germanium detector under study with collimator.}
\label{fig:ClusterSize}
\end{figure}

The distribution of the cluster size for the germanium detector with collimator is presented in Figure~\ref{fig:ClusterSize} (a). The cluster size multiplicity shows that the large majority of events have a cluster size equal to 1 (these are events collected in a single pixel), while less than 2.8\% of the events are contributing to the charge sharing effect. This result illustrates the important role of the collimator in removing the charge sharing effect. The 2D map in x and y, depicted in Figure~\ref{fig:ClusterSize} (b), provides more details on the spatial distribution of the cluster size. The largest cluster size occurs in the inter-pixel region, as expected, since the low electric field present in this area results in a strong charge carrier diffusion. As a consequence, the induced charge carriers of one pixel can easily diffuse into adjacent pixel cells, resulting in a cluster size of 2 to 4 in the corner. On the other hand, the cluster size is 1 at the center of the pixel because charge carriers are unlikely to diffuse to neighboring pixel cells.

\begin{figure}[htb!]
\centering
\subfloat[]{\includegraphics[width=0.40\textwidth]{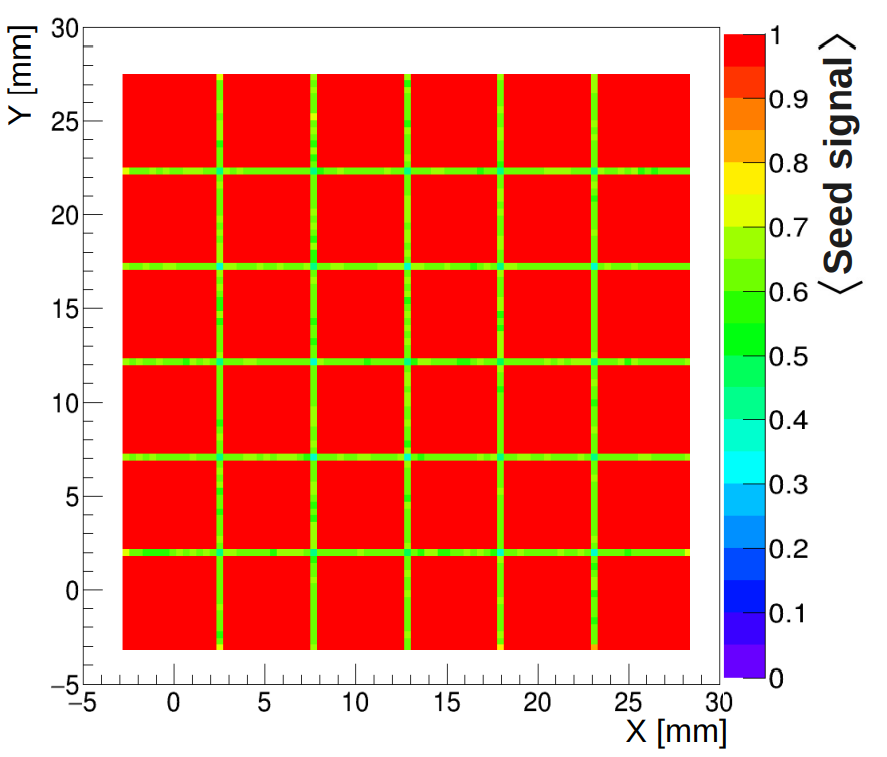}}
\qquad
\subfloat[]{\includegraphics[width=0.40\textwidth]{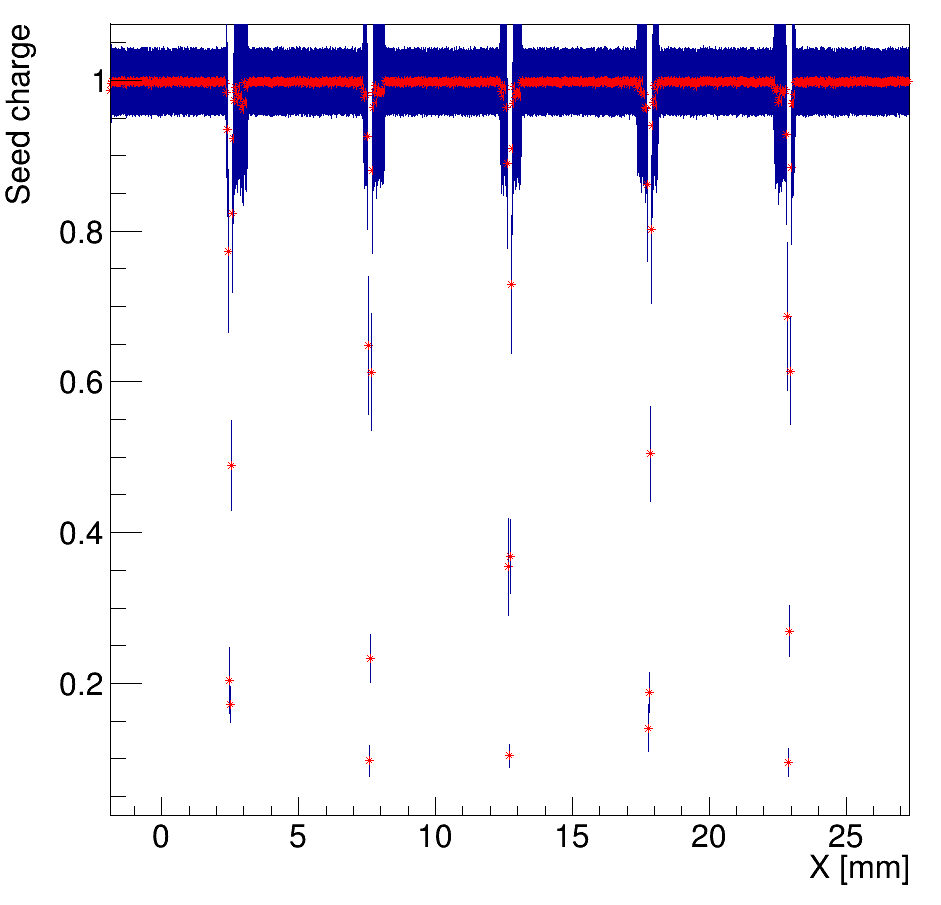}}

\caption{(a) Seed signal map and (b) seed signal distribution along a horizontal cut-line crossing the middle of a pixel row for the multi-element germanium detector under study with collimator.}
\label{fig:SeedSginal}
\end{figure}

Figure~\ref{fig:SeedSginal} (a) shows the seed signal map in x and y and depicts a decrease of the seed signal towards the edges due the presence of the collimator.
A scan of the seed signal along a horizontal cut-line crossing the middle of a pixel row is shown in Figure~\ref{fig:SeedSginal} (b) in which the seed signal drops down to 10\% of the maximum in the inter-pixel region, while the vertical lines represent the error bars for each data point.

\begin{figure}[htb!]
\centering
\subfloat[]{\includegraphics[width=0.4\textwidth]{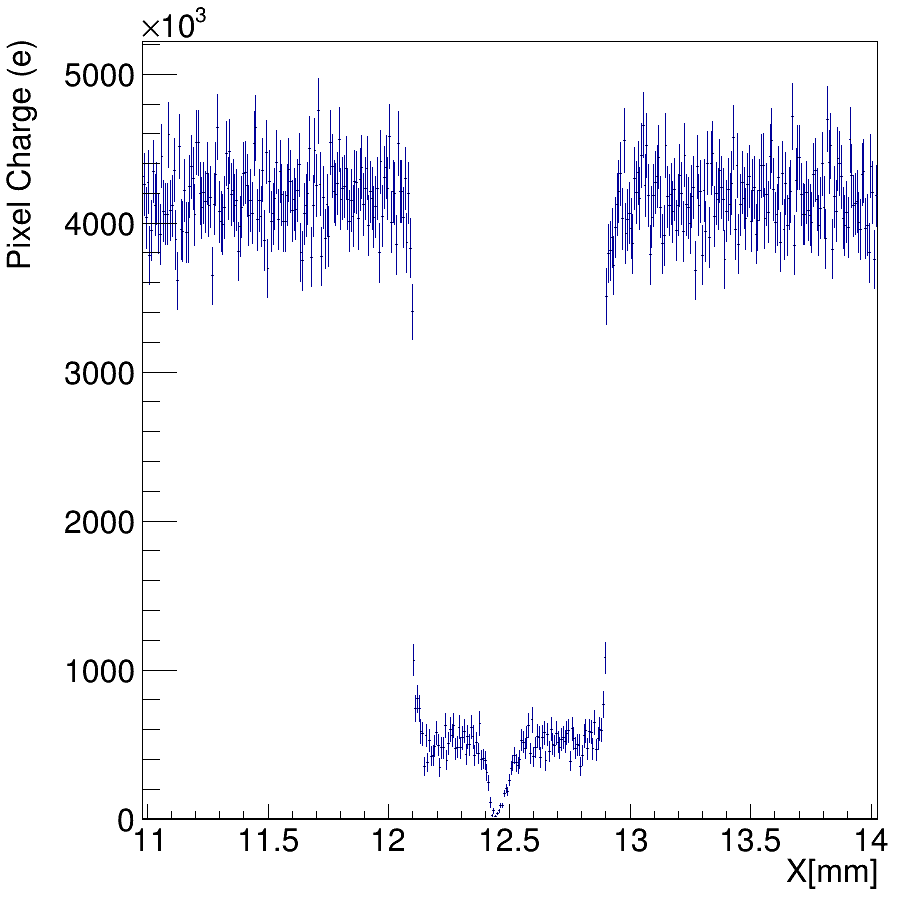}}
\qquad
\subfloat[]{\includegraphics[width=0.4\textwidth]{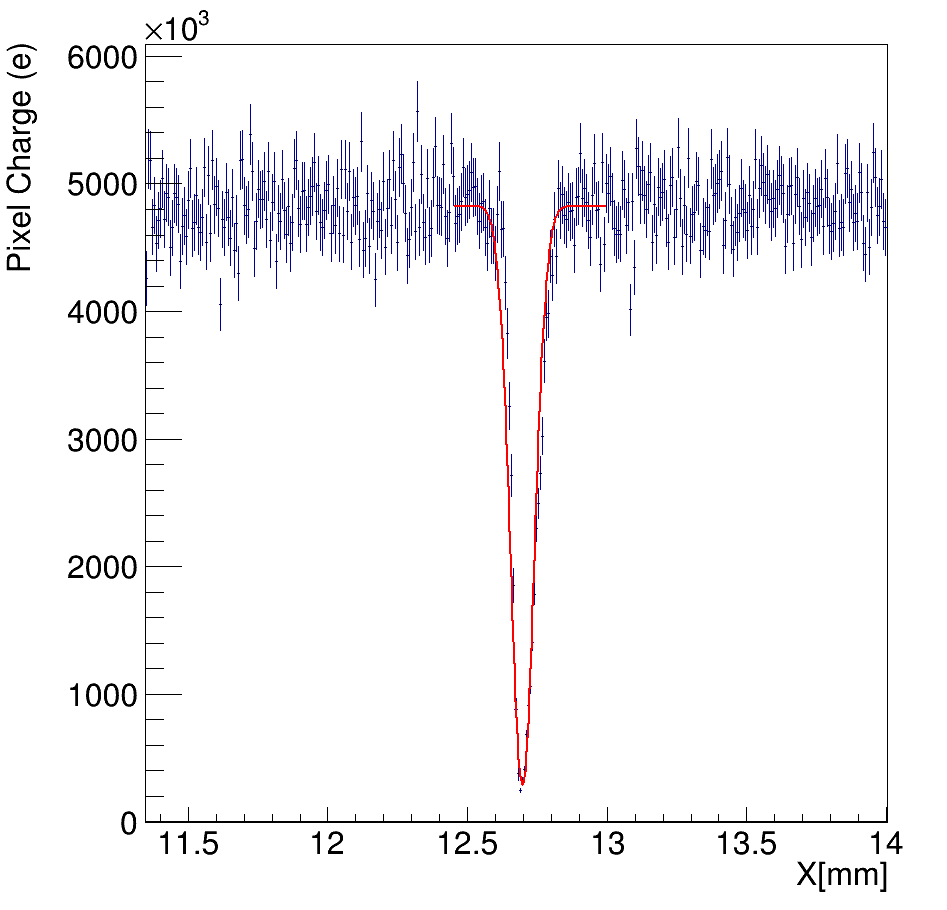}}

\caption{Pixel charge scan, along a horizontal cut-line crossing the middle of a pixel row, for the multi-element germanium detector under study with collimator (a), without collimator (b).}
\label{fig:PixelCharge}
\end{figure}

The charge sharing effect can also be quantified by measuring the extent of the charge sharing region, defined as the region where the signals from two neighbor pixels are below 90\% on both sides, i.e. the charge is shared. This width is calculated from Figure~\ref{fig:PixelCharge}, where a pixel charge scan is done between two adjacent pixels. In the case where there is no collimator (Figure~\ref{fig:PixelCharge} (b)), a Gaussian fit is used to estimate the charge sharing region extent. In both cases, the pixel charge scan shows that in the inter-pixel region, the pixel charge drops down to about 10\% - 20\% of the total signal. The charge sharing region width is estimated to be 114~\textmu m when removing the collimator, which is much smaller than the inter-pixel gap of 800~\textmu m. When the collimator is present, there is a 1~mm wide area between pixels with a low efficiency, i.e., pixels are well isolated but at the same time, there is a significant reduction of the active area of about 30\%. In terms of charge sharing events, i.e. those events with a cluster size greater than 1, there is an increase from 3\% to 8\% if the collimator is removed, as shown in Table~\ref{tab:DistanceChargeSharing}. Hence, removing the collimator can be a good strategy to increase the active area in some experiments. The comparison in detector performance is enlarged to a wide range of energies in the following section.

\begin{table}[t]
\begin{threeparttable}
\caption{List of the different observables used in this study of charge sharing effect and the estimated values by simulation for a multi-element germanium detector equipped with and without collimator.}
\label{tab:DistanceChargeSharing}
\begin{tabular}{p{8cm}cc}
\hline
Parameter & With collimator & Without collimator \\
\hline
Pixel size [$mm^{2}$] & $4.2\times4.2$ & $4.2\times4.2$  \\
Detector Sensitive Area [$mm^{2}$] & 635 & 820 \\
Sensor thickness [mm] & 7 & 7  \\
Shared events [norm. \%]& 3\% & 8\% \\
Signal efficiency\tnote{a} (\%) & 95\% & 88\%\\
Background efficiency\tnote{b} (\%) & 0.04\% & 0.50\% \\
Signal-to-Background ratio & 2375 & 180\\
Charge sharing width\tnote{c} (\textmu m) & 1003 &114 \\
\hline
\end{tabular}
\begin{tablenotes}
\item [a] The signal efficiency is the number of hits in the signal RoI divided by the total number of hits.
\item [b] The background efficiency is the number of hits in the background RoI divided by the total number of hits.
\item [c] In case of the detector equipped with collimator, the \textit{charge sharing width} is equivalent to a \textit{low efficiency region width}.
\end{tablenotes}
\end{threeparttable}
\end{table}

\subsection{Study of the detector performance at different beam energies}
Further simulations have been performed to study the performance of the germanium detector under study with and without titanium collimator at different beam energies between 5 and 80 keV. As shown in the previous section, a uniform photon beam perpendicular to the detector frontside has been considered. In this section, the different simulation results are presented, in particular the number of shared events and the signal-to-background ratio (S/B).

\begin{figure}[htb!]
\centering
\includegraphics[width=0.6\textwidth]{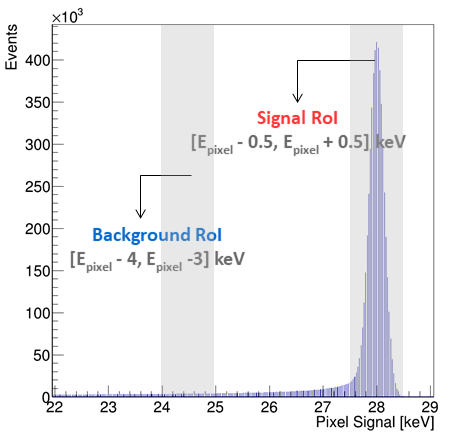}
\caption{Energy spectrum for a photon beam of energy 28~keV shows the signal and background RoI regions defined for this study. Shaded areas define the boundaries of the different RoI regions.}
\label{fig:ROI_PixelSignal}
\end{figure}

The S/B ratio is defined as the number of events in a signal RoI (Region of Interest) divided by the number of events in the background RoI as illustrated in Figure~\ref{fig:ROI_PixelSignal}, where an incident photon beam energy of 28~keV is used in this example. In the energy spectrum, the signal RoI is defined at the position of the photoelectric peak in the range [E$_{photo}$ - 0.5, E$_{photo}$ + 0.5] keV, while the background RoI is defined in the range [E$_{photo}$ - 4, E$_{photo}$ -3] keV, an energy region in the Compton level close to the signal RoI. This choice of the background RoI is an estimation of the background level in the signal RoI. For this study, all pixels with energy less than 1 keV have been excluded since all the digital electronics, e.g. X-MAP, XIA-FalconX and Xspress3X that are used during experiments on the beamline, currently apply this cut on energy. Therefore, this selection has been applied in order to have a better agreement later on between data and simulation.

\begin{figure}[!t]
\centering
\subfloat[]{\includegraphics[width=0.45\textwidth]{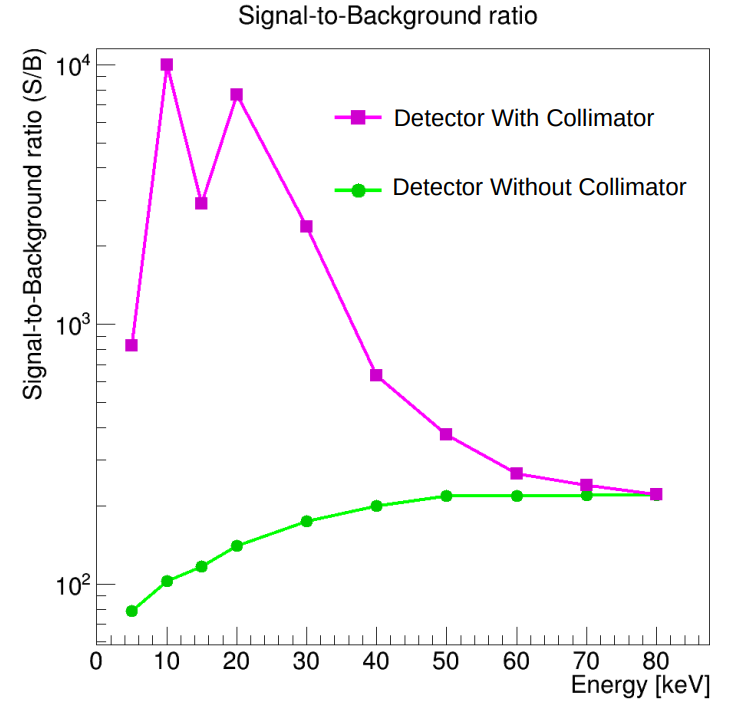}}
\qquad
\subfloat[]{\includegraphics[width=0.46\textwidth]{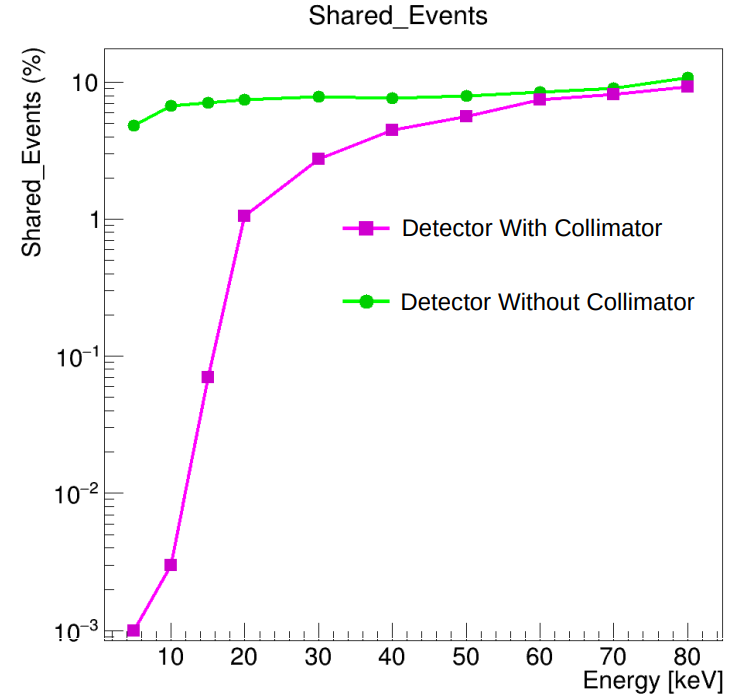}}

\caption{The signal-to-background (S/B) ratio (a) and the number of charge sharing events (b) as a function of the incident photon beam energy for a multi-element germanium detector with and without collimator.}
\label{fig:SB_Shared}
\end{figure}

Figure \ref{fig:SB_Shared} (a) shows the S/B ratio as a function of the beam energy. This result shows that the S/B ratio is strongly dependent on the incident beam energy for the with-collimator configuration, while the dependence is smoother in absence of a collimator.
The S/B ratio is higher for a with-collimator design and for beam energies lower than 60~keV. For energies higher than 60~keV, the S/B ratio for both configurations is nearly the same showing that the collimator has no longer a significant effect at high energy and becomes transparent to incident X-rays. For the with-collimator configuration, an abrupt drop of the S/B ratio is observed between 10 and 15~keV, due to the loss of signal efficiency related to the presence of $K_\alpha$ and $K_\beta$ emissions lines of germanium, as already observed in other studies \cite{deloule:cea-01791938}. 

The number of shared events as a function of beam energy is shown in Figure \ref{fig:SB_Shared} (b). This study shows that the number of shared events increases for higher beam energies in both configurations. At energies lower than 20~keV, the collimator helps in removing almost completely all shared events. In the configuration without collimator, the number of shared events shows a plateau in the energy range 30 and 60~keV, as charge sharing is mainly determined by \textit{split events} and this number does not change with energy. In fact, the size of charge cloud increases with energy but photons are also absorbed deeper and closer to the pixel implant, which countervail the other effect~\cite{BARYLAK2018234}. Meanwhile, at energies higher than 60~keV, the number of shared events increases as it is mainly determined by \textit{fluorescence events}, i.e., photons absorbed in one pixel and one escaping fluorescence photon absorbed in another pixel. In both cases, the number of shared events increases up to 10\% at 80~keV.

The results herein obtained shows that the simulation chain proposed in this paper is fully operational and allowed us to quantify some interesting performance parameters of the multi-element germanium detector under study. Further studies and detector design improvement can benefit from using these simulation tools.

\section{Measurements to validate simulation results}
\label{sec:validation}
In order to setup and validate the simulation chain proposed in this paper, a comparison between simulation and experimental data in two particular cases has been performed and is presented in this section. The experimental data was recorded at the SAMBA beamline of SOLEIL synchrotron using a multi-element germanium detector under study, previously described in Section~\ref{sec:detmodel}. During the on-beamline measurements, an energy spectrum of 60~seconds acquisition time has been recorded for each of the 36~active pixels of the detector and using the XIA-DXP-xMAP DPP.

\subsection{On-beamline measurements}
\label{sec:expsetup}
\begin{figure}[htb!]
\centering
\subfloat[]{\includegraphics[width=0.45\textwidth]{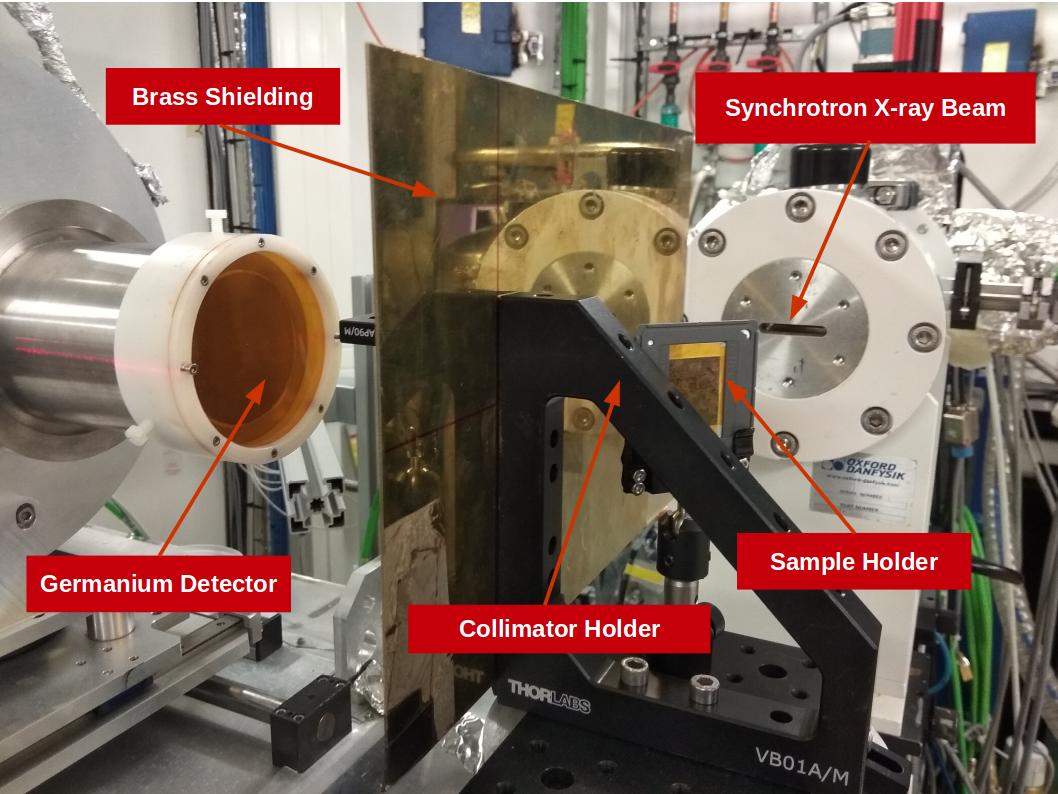}}
\qquad
\subfloat[]{\includegraphics[width=0.45\textwidth]{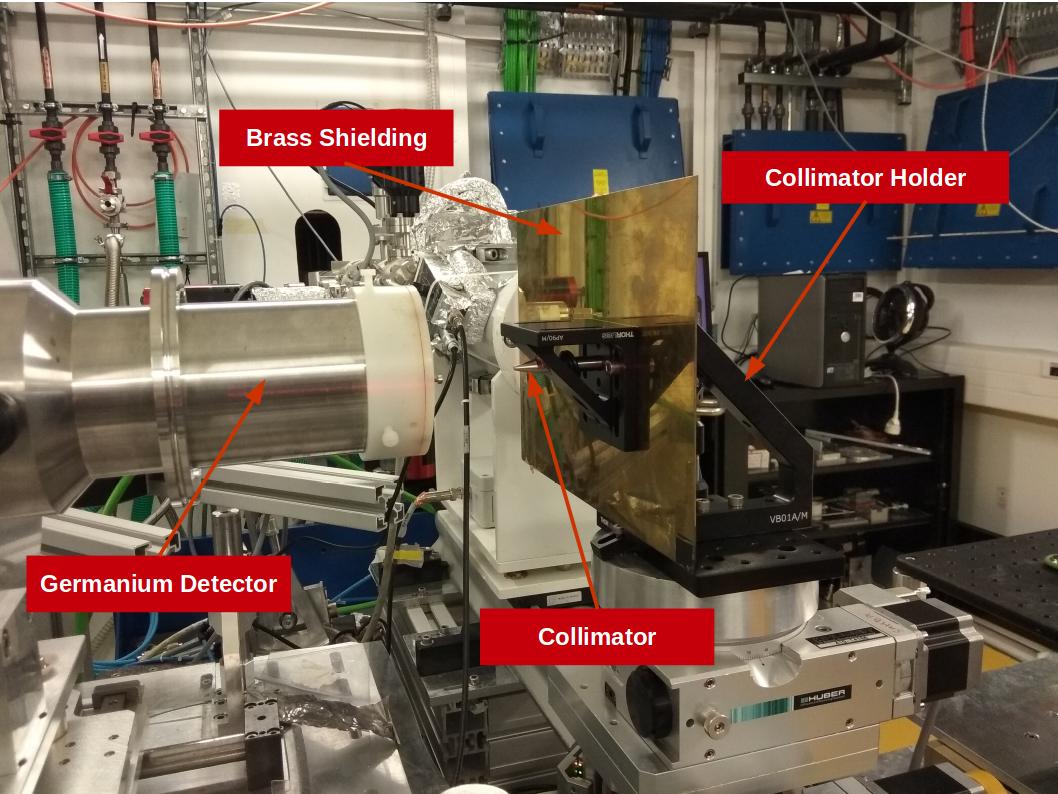}}
\caption{(a) Experimental setup used for the charge sharing effect investigation at the SAMBA beamline and (b) another view of the setup shows the collimator placed between the germanium detector and the sample holder.}
\label{fig:SAMBACollimator}
\end{figure}

Two different measurements have been performed at the SAMBA beamline. The first one, illustrated in Figure~\ref{fig:SAMBATarget} (a) and (b), was a fluorescence measurement of a complex sample, with the aim at reproducing with our simulation all the X-ray lines emitted by this sample. In the second measurement, illustrated in Figure~\ref{fig:SAMBACollimator} (a) and (b), the objective was to study the detector performance in the inter-pixel region by performing a horizontal scan using a collimated beam and to compare experimental results to simulation ones. The two setups are described below:

\begin{itemize}
\vspace{-0.2cm}
    \item \textbf{Experimental measurement with a Cadmium sample and wide detector illumination:}
    for this measurement, a reference sample composed of cadmium (0.918\%), iron (0.26\%), and traces of lead (20~ppm), strontium (30~ppm), zirconium (20~ppm), and copper (3~ppm) has been used. While the rest of the sample is organic soil, i.e., a mixture of carbon, hydrogen and oxygen elements. The energy of the beam was set to 28.0~keV by a monochromator situated upstream of the beamline. The beam size was then focused in a small area at the sample spot (385~\textmu m $\times$ 250~\textmu m widths in the horizontal and vertical axis) and a photon flux of~$3.47 \times 10^{10}$ ph/sec was measured by the ionization chamber situated just before the sample. Then, a cylindrical sample (10~mm diameter, 1.0~mm thick, a density of 1.28~g/cm$^3$) was placed at the sample spot, at an angle of 45~deg between the beam and the detector, and the germanium detector was situated at a sample-to-detector distance of 312~mm, corresponding to an input count rate of $\sim$28~kcps and a dead time of $\sim$13\% measured for all pixels. In this case, this dead time value is small enough to keep a linear relationship between an element concentration and the sum of measured X-rays fluorescence intensities at germanium detector pixels. The setup of this measurement is shown in Figure~\ref{fig:SAMBATarget} (a) and (b).

\begin{figure}[htb!]
\centering
\subfloat[]{\includegraphics[width=60mm]{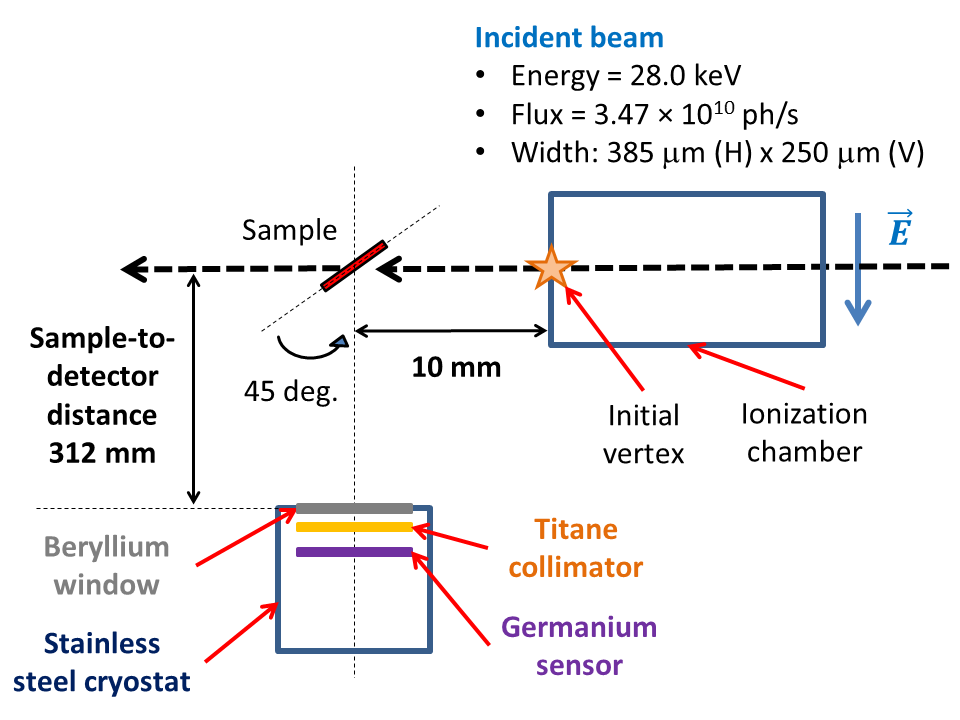}}
\subfloat[]{\includegraphics[width=90mm]{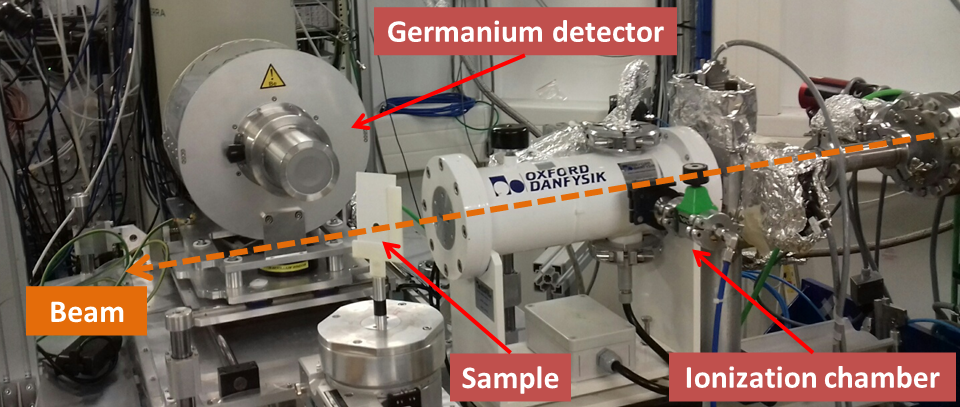}}
\caption{(a) General layout of a typical on-beamline measurement at the SAMBA beamline. (b) Picture of the whole experimental setup used for fluorescence measurement.}
\label{fig:SAMBATarget}
\end{figure}

\item \textbf{Experimental measurement with a tin foil and collimated detector illumination:}
this second measurement aimed at studying the charge sharing effect in the inter-pixel region between two adjacent pixels. To achieve this objective, the previous experimental setup has been modified, as shown in Figure~\ref{fig:SAMBACollimator} (a). In this setup, a collimator of 1~mm diameter, 6~cm length and made of tungsten and stainless steel was used. The collimator was placed between the germanium detector and the sample holder at 5~mm distance from the detector window (see Figure~\ref{fig:SAMBACollimator} (b)). This piece allowed collimating the fluorescence emitted by a tin foil, situated at the sample spot and at 45~deg between the beam and the detector. In addition, a plate of brass was used to shield diffused X-rays from the setup. The energy of the beam was then set to 30.8~keV and the beam spot was focused in a small area of the tin foil (308~\textmu m $\times$ 300~\textmu m widths in the horizontal and vertical axis). Before the measurement, the different components (i.e. beam spot, collimator and detector) were aligned. Several horizontal scans of one row composed of 6 pixels have been made with a step size of 0.2~mm.
\end{itemize}

\subsection{Comparison between simulation and experimental results}
\label{sec:energyspecta}

For the first experiment with the cadmium sample, a comparison of measured and simulated energy spectra is shown in Figure~\ref{fig:EnergyDistTarget}.
A fair agreement between data and simulation is observed,
especially for the main X-ray lines of the chemical elements cadmium, iron, strontium and zirconium, and also the escape peaks of cadmium X-ray emission lines.
Despite the good agreement for the X-ray emission lines of the elements,
we note two differences between experimental and simulated spectra:

\begin{figure}[htb!]
\centering
\includegraphics[width=0.9\textwidth, height = 0.6\textwidth]{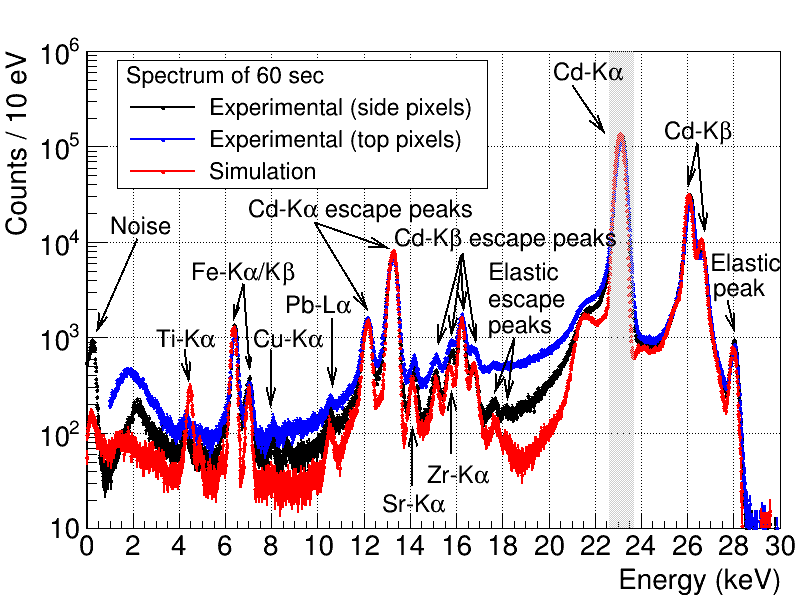}
\caption{Simulated and measured energy spectrum from the on-beamline measurement with a reference sample rich in cadmium at the SAMBA beamline. The simulated energy spectrum (red) is compared to two experimental spectra measured on two different areas of the matrix to show the dependence of the Compton level on the pixel's position. The shaded grey area defines the RoI of the K$\alpha$ line of Cadmium.}
\label{fig:EnergyDistTarget}
\end{figure}

\begin{itemize}
    \item The Compton level created by cadmium fluorescence lines (energies between 2 and 21~keV), is higher in experimental data. This difference is due to the fact that experimental data includes an extra contribution by the preamplifier reset signals, which has not been yet modeled in our simulation and affects the whole spectrum. Moreover, the detector used in these measurements has two disconnected (noisy) pixels, while our simulation assumes a fully functional detector. As a result, the experimental Compton level seems correlated to the distance to these two disconnected pixels, situated at the top part of the germanium sensor.
    This dispersion of Compton level along the sensor surface is illustrated by the energy spectrum of 6 pixels around the disconnected ones (blue curve of Figure~\ref{fig:EnergyDistTarget}) and that of side pixels (black curve). The side pixels spectrum shows a better agreement with the simulated spectrum of a fully functional detector.
    
    \item The presence of titanium fluorescence X-ray emission lines (at 4.51 and 4.93~keV) in the simulation, while they are absent in experimental spectrum. The titanium collimator is probably coated with aluminium to block its fluorescence lines as cited here~\cite{Chatterji2016}. This coating has not been included in the detector model due to the absence of information in the detector datasheet.
\end{itemize}

\begin{figure}[htb!]
\centering
\includegraphics[width=0.9\textwidth, height = 0.60\textwidth]{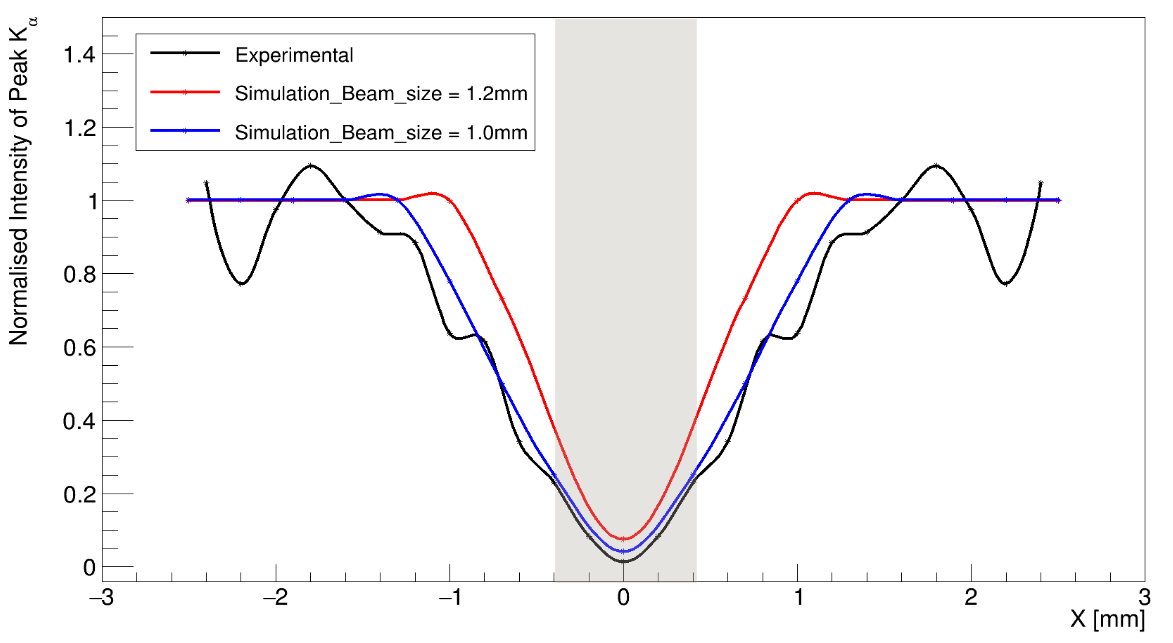}
\caption{The intensity of the K$_{\alpha}$ line of tin as a function of horizontal scan position. Comparison between simulation and on-beamline measurement with Sn foil at the SAMBA beamline. The shaded area define the limits of the inter-pixel region.}
\label{fig:Intensity}
\end{figure}

Regarding the second experimental measurement, the multi-element germanium detector performance in the inter-pixel region has been investigated. Thanks to a collimated beam emitted from the tin foil, the intensity of the K$_{\alpha}$ line of tin (i.e. the total number of counts in the RoI [24.7 keV - 25.7 keV]) has been studied as a function of the horizontal position and shown in Figure~\ref{fig:Intensity}.

This figure shows a comparison of the normalized intensity of the K$_{\alpha}$ line for experimental data and two simulation cases, corresponding to two different beam diameters (1.0 and 1.2~mm). The best agreement between experimental and simulation is obtained for a beam diameter of 1.0~mm, which is coherent with the collimator diameter of 1~mm used in this measurement. The intensity of the K$_{\alpha}$ line increases slightly for a larger beam size of 1.2~mm. This can be explained by the fact that as the beam spot is getting larger, it is covering a larger area. In the inter-pixel region, the number of counts (hits) is negligible ($\approx$ 5\% of the maximum intensity) and a good agreement between simulation and measurement is observed.
We can also notice some oscillation in the experimental data that could be explained by the non-uniformity of the incident beam (unfortunately not measured during the experiment) and the non perfect stability of our detector mechanical motorized table after each scan step. Despite those small differences, we conclude that our simulation chain is fully operational and is validated.


\section{Conclusion and Outlook}
\label{sec:con}
This work presents a first complete and operational simulation chain based on Allpix Squared framework, customized to multi-element germanium detectors and combined with three-dimensional simulations of the electric field and the weighting potential, based on COMSOL Multiphysics®. Using this simulation chain, a quantification of charge sharing effect as well as signal-to-background ratio has been made for X-ray beams in an energy range from 5 to 80~keV and for a germanium detector equipped with and without collimator.

For a beam energy of 30~keV, charge sharing events are 3\% if the detector is equipped with a collimator and 8\% in absence of it. These events are collected in the inter-pixel region, as expected, since the low electric field results in a strong charge carrier diffusion. In absence of collimator, the charge sharing region width found by simulations is about~114~\textmu m, while the detector sensitive area in increased by about 30\%.

The signal-to-background (S/B) ratio shows a strong dependence on the beam energy for a germanium detector equipped with a collimator, while this dependence is smoother in absence of it. The first configuration shows higher values for S/B ratio for beam energies lower than 60~keV. At higher beam energies, there is no difference between the two configurations because the collimator becomes transparent. In terms of charge sharing events, the collimator helps in removing almost completely all charge sharing events for beam energies lower than 20~keV, while this number increases up to 10\% at 80~keV, the same value as that of the configuration without collimator.

Finally, two experimental measurements have been performed at the SAMBA beamline at SOLEIL synchrotron. The experimental data were used to set up the full simulation chain and good agreements have been observed between data and simulation.

This simulation chain provides a powerful tool that could drive current and future developments of new generations of multi-element germanium detectors. Further studies to investigate the detector performance in different configurations are ongoing, i.e. for smaller pixel sizes and different pixels geometries. These studies will aim at optimizing the detector design, in order to enhance the S/B ratio for detecting chemical species with very low concentrations (less than 10~ppm). Scientific cases of XAFS experiments could also be studied, such as the detection of heavy element traces, like cadmium or lead, over a background level created by beam X-rays (and commonly known as elastic peak); or of light elements in a complex sample, whose background level is mainly defined by the Compton level of more intense fluorescence lines situated at higher energy.

\acknowledgments
The authors acknowledge E. Fonda and G. Landrot, both scientists of the SAMBA beamline at SOLEIL synchrotron, for their cooperation in providing beamtime and information about the samples used during the two measurements performed at the SAMBA beamline in 2020 and 2021. The authors would like also to thank SOLEIL computing service (ISI), in particular to Ph. Martinez, for technical support in the use of SUMO and TGCC COBALT IT clusters.

\bibliographystyle{JHEP}
\bibliography{mybibfile}
\end{document}